\documentclass[aps,twocolumn,pra,superscriptaddress,amsmath,amssymb]{revtex4-2}
\usepackage{dcolumn}
\usepackage{bm}
\usepackage{amsmath}
\usepackage{txfonts}
\usepackage[T1]{fontenc}
\usepackage{xspace}
\usepackage{ulem}
\usepackage{comment}
\usepackage{braket}
\ifx\pdfoutput\undefined
\usepackage[dvipdfmx]{graphicx}
\usepackage[dvipdfmx]{hyperref}
\usepackage[dvipdfmx]{color}
\else
\usepackage{graphicx}
\usepackage{hyperref}
\usepackage{color}
\fi

\begin{document}
\let\emph\textit

\title{
  Modulated honeycomb lattices and their magnetic properties
}

\author{Akihisa Koga}
\affiliation{
  Department of Physics, Institute of Science Tokyo,
  Meguro, Tokyo 152-8551, Japan
}

\author{Toranosuke Matsubara}
\affiliation{
  Department of Physics, Institute of Science Tokyo,
  Meguro, Tokyo 152-8551, Japan
}

\date{\today}
\begin{abstract}
  We propose a family of modulated honeycomb lattices, 
  a class of quasiperiodic tilings characterized by the metallic mean.
  These lattices consist of six distinct hexagonal prototiles
  with two edge lengths, $\ell$ and $s$, and can be regarded as
  a continuous deformation of the honeycomb lattice.
  The structural properties are examined
 through their substitution rules. 
  To study the electronic properties,
  we construct a tight-binding model on the tilings,
  introducing two types of hopping integrals, $t_L$ and $t_S$,
  corresponding to the two edge lengths, $\ell$ and $s$, respectively.
  By diagonalizing the Hamiltonian on these quasiperiodic tilings, 
  we compute the corresponding density of states (DOS). 
  Our analysis reveals that
  the introduction of quasiperiodicity in the distribution of hopping integrals
  induces a spiky structure in the DOS at higher energies, while
  the linear DOS at low energies ($E\sim 0$) remains robust. 
  This contrasts with the smooth DOS in the disordered tight-binding model,
  where two types of hopping integrals are randomly distributed
  according to a given ratio.
  Furthermore, we study the magnetic properties of the Hubbard model
  on modulated honeycomb lattices
  by means of real-space Hartree approximations.
  A magnetic phase transition occurs at a finite interaction strength
  due to the absence of the noninteracting DOS at the Fermi level.
  When $t_L\sim t_S$, the phase transition point is primarily governed by
  the linear DOS.
  However, far from the condition $t_L=t_S$,
  the quasiperiodic structure plays a significant role
  in reducing the critical interaction strength,
  which is in contrast to the disordered system. 
  Using perpendicular space analysis,
  we demonstrate that sublattice asymmetry
  inherent in the quasiperiodic tilings
  emerges in the magnetic profile,
  providing insights into the interplay
  between quasiperiodicity and electronic correlations.
\end{abstract}

\maketitle

\section{Introduction}

Quasicrystals~\cite{Shechtman_1984,Levine} provide an essential bridge
between periodic and disordered systems.
Unlike conventional crystals, which exhibit translational symmetry, and
amorphous materials, which lack long-range order,
quasicrystals possess an aperiodic yet long-range ordered structure.
This dual nature enables quasicrystals
to combine elements of both order and complexity,
making them a subject of significant scientific interest.
Understanding these properties requires a comprehensive examination of 
their structural characteristics and electronic properties. 
Consequently, a unified framework is needed for describing periodicity,
quasiperiodicity, and disorder on an equal footing.

Toy models defined by point-to-point connectivity,
such as the tight-binding, Hubbard, and Heisenberg models,
serve as potential candidates for simultaneously incorporating
periodicity, quasiperiodicity and disorder.
The effect of disorder can be studied by introducing randomness
in the intersite couplings.
However, incorporating both periodicity and quasiperiodicity
remains challenging due to the unique rotational symmetry of quasicrystals,
which is forbidden in periodic systems.
This limits research in this field to simple cases,
such as the one-dimensional Fibonacci chain~\cite{Kohmoto_1983,Kohmoto_1986,Kohmoto_1987,Fujiwara_1989,Mace_2017,Jagannathan_2021,Iijima_2022} and the square Fibonacci lattice~\cite{sFib,Even_2006,Even_2008}.

In our previous study~\cite{Matsubara}, we have proposed
quasiperiodic hexagonal tilings,
using the Fibonacci sequence.
These tilings can be regarded as the continuously deformed versions of
the triangular, dice, and honeycomb lattices.
This allows us to discuss the effect of the quasiperiodicity
on the regular honeycomb lattice.
Furthermore, examining the theoretical models on the tilings,
we can clarify
how the quasiperiodicity and disorder
affect intriguing low-energy properties
of the strongly correlated electron systems. 
One of the important questions is how robust Dirac-type dispersions
and van-Hove singularity in the honeycomb system,
which should play an important role in stabilizing
the spontaneously symmetry-breaking state~\cite{Sorella_1992,Furukawa_2001,Feldner_2010,Sorella_2012,Assaad_2013,Raczkowski_2020,Ostmeyer_2020,Ostmeyer_2021,Takemori},
remain under quasiperiodicity and disorder.
Therefore, modulated honeycomb lattices provide
an appropriate platform to systematically discuss the effects of
periodicity, quasiperiodicity and disorder.

The honeycomb lattice structure modulated by the golden mean
has been proposed in our previous work~\cite{Matsubara},
while the structures of generic metallic-mean tilings remain unexplored.
The substitution rule, which has not yet been given,
is expected to be crucial
for systematically examining the toy models.
Here, we propose a set of substitution rules
to generate modulated honeycomb lattices
associated with the generic metallic mean.
These tilings are composed of six types of prototiles, and
their tile and vertex properties are analyzed based on these rules.
To investigate the effects of quasiperiodicity and disorder
in the electron systems on the honeycomb lattice,
we examine the tight-binding models,
focusing on their density of states (DOS).
Furthermore, the magnetic properties are discussed
by applying site-dependent real-space Hartree approximations
to the Hubbard model.
We find that a magnetic phase transition occurs
at a finite interaction strength
due to the absence of the noninteracting DOS at the Fermi level.
The magnetic profile inherent in quasiperiodic tilings
is also addressed within perpendicular space.

The paper is organized as follows. 
In Sec.~\ref{sec:lattice}, 
we introduce the modulated honeycomb lattices,
and clarify their tile and vertex properties. 
We construct the tight-binding model on these tilings
and examine their DOS in Sec.~\ref{sec:tb}. 
By means of the site-dependent real-space Hartree approximations, 
we clarify how a magnetically ordered state competes with a semimetallic state 
in the Hubbard model in Sec.~\ref{sec:mag}. 
A summary is given in the last section.

\section{Modulated honeycomb lattice}\label{sec:lattice}
\begin{figure*}[htb]
  \includegraphics[width=\linewidth]{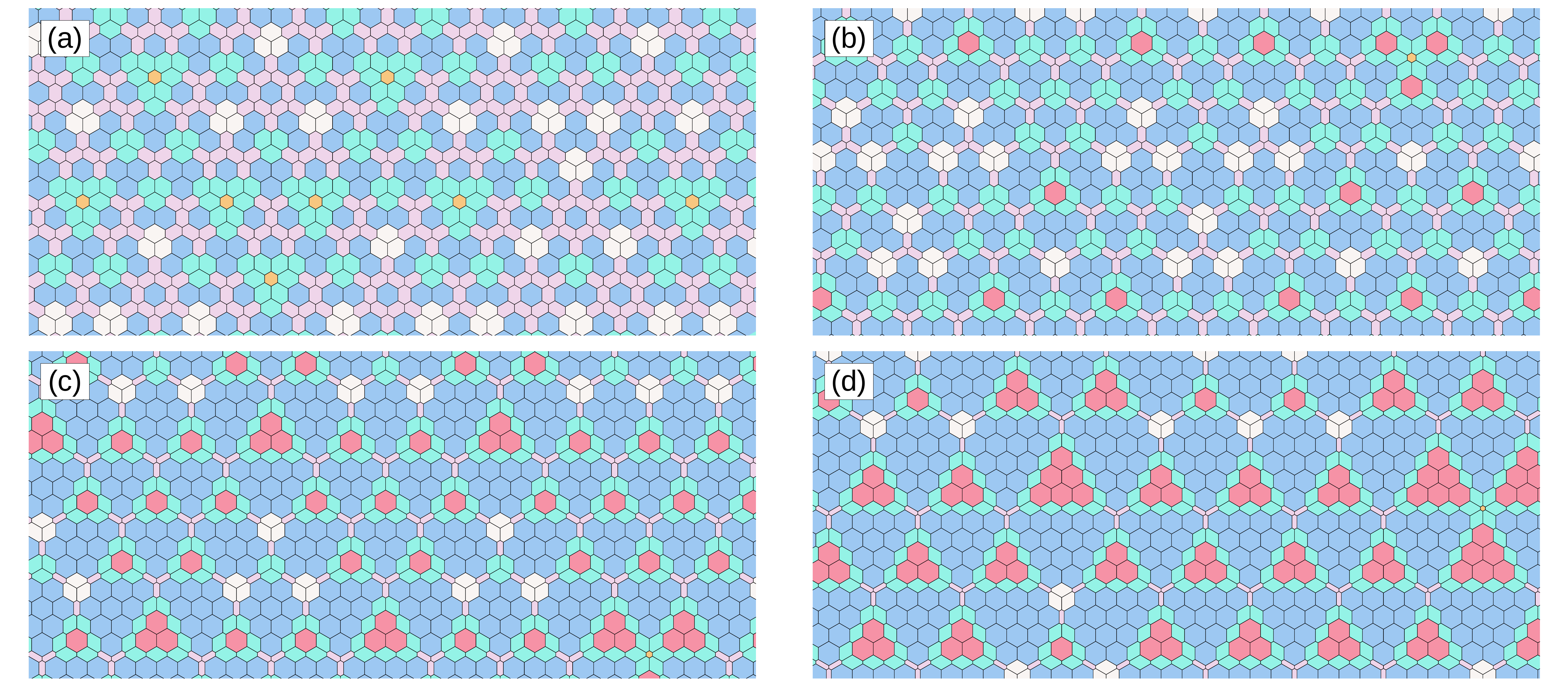}
  \caption{
    Modulated honeycomb lattices characteristic
    of (a) golden-mean, (b) silver-mean, (c) bronze-mean, and (d) 4th metallic-mean. 
  }
  \label{lattice}
\end{figure*}

In this study, we consider the modulated hexagonal tilings
characteristic of the metallic mean $\tau_k=(k+\sqrt{k^2+4})/2$,
as shown in Fig.~\ref{lattice}.
Since the tilings are constructed by densely packing
the two-dimensional plane with hexagons whose inner angles are $2\pi/3$,
they should be regarded as continuously deformed honeycomb lattices.
This suggests that the quasiperiodic structure can be introduced
into the toy models defined on the honeycomb lattice,
which will be discussed in the following sections.

\begin{figure}[htb]
  \includegraphics[width=0.8\linewidth]{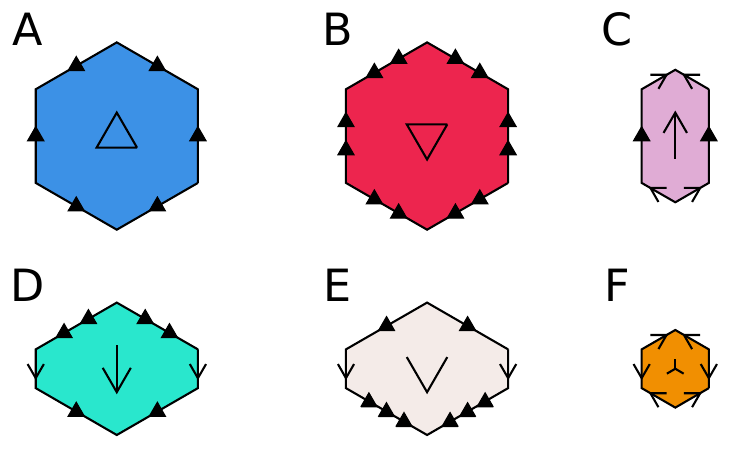}
  \caption{
    Six types of hexagonal tiles for the modulated honeycomb lattices,
    characterized by the metallic mean. 
    The ratio between long and short lengths is $\tau_k$.
    The matching rule of the directed tiles is indicated 
    by the solid triangles and arrows on their edges.
    Tiles A, B, C, D, E, and F are distinguished by 
    the symbols located at their centers. 
  }
  \label{6tiles}
\end{figure}

The details of the tilings are explained in this section.
The prototiles in these tilings consist of six types of directed hexagons,
as shown in Fig.~\ref{6tiles}.
The A and B (F) tiles are large (small) regular hexagons
with edge length $\ell$ ($s$),
where $\ell=\tau_k s$.
The C (D and E) tile is a thin (fat) hexagon with lengths $\ell$ and $s$.
The matching rules for the directed tiles are indicated by 
the solid triangles and arrows on their edges in Fig.~\ref{6tiles}.

\begin{figure}[htb]
  \includegraphics[width=\linewidth]{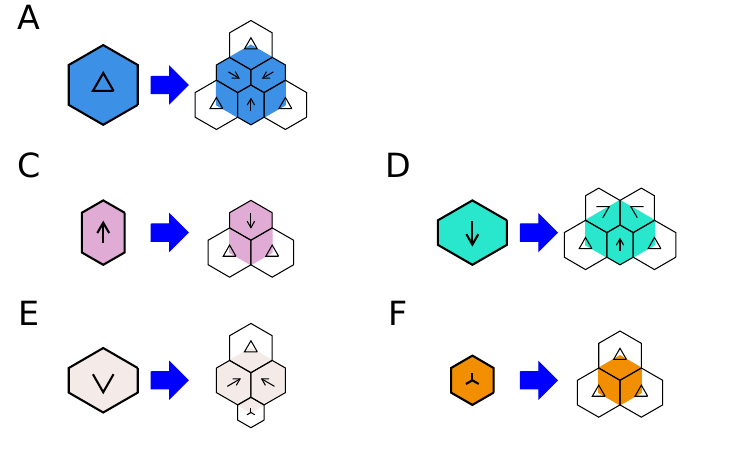}
  \caption{
    Substitution rule for the golden-mean modulated honeycomb lattice.
  }
  \label{gold}
\end{figure}

\begin{figure}[htb]
  \includegraphics[width=\linewidth]{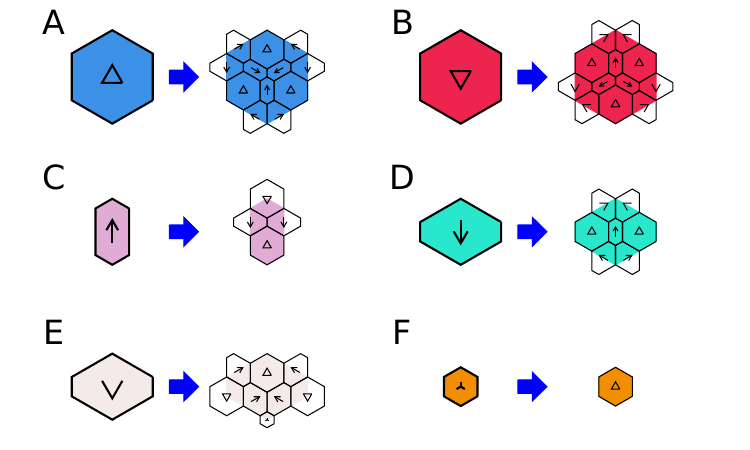}
  \caption{
    Substitution rule for the silver-mean modulated honeycomb lattice.
  }
  \label{silver}
\end{figure}

\begin{figure}[htb]
  \includegraphics[width=\linewidth]{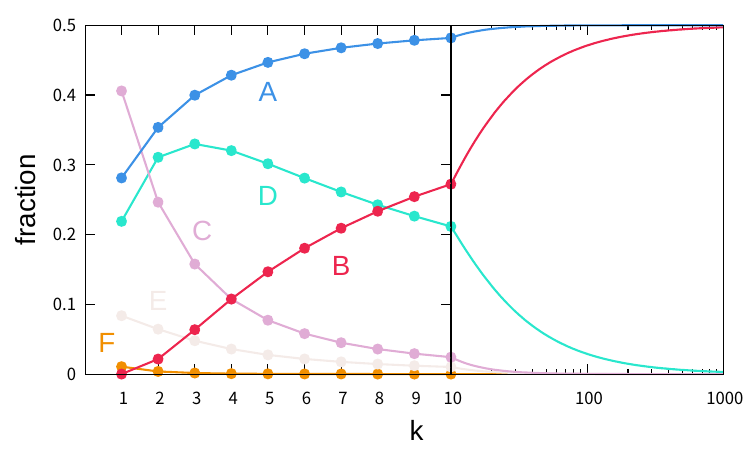}
  \caption{
    Tile fractions of the $k$th metallic-mean modulated honeycomb lattices.
  }
  \label{frac}
\end{figure}

The modulated honeycomb lattices are generated,
using substitution rules.
The substitution rules for golden-mean ($k=1$) and silver-mean ($k=2$) tilings
are explicitly presented in Fig.~\ref{gold} and Fig.~\ref{silver}.
Although the golden-mean tiling is an exception in this series
due to the lack of B tiles,
we develop substitution rules for the generic metallic-mean tilings.
The details are provided in Appendix~\ref{sub}.
The number of six types of tiles increases under the substitution operation 
for the $k$th metallic-mean tiling as
${\bf v}_k^{(n+1)}=M_k{\bf v}_k^{(n)}$, where
${{\bf v}_k^{(n)}}=(N_{k{\rm A}}^{(n)}, N_{k{\rm B}}^{(n)}, N_{k{\rm C}}^{(n)}, N_{k{\rm D}}^{(n)}, N_{k{\rm E}}^{(n)}, N_{k{\rm F}}^{(n)})^t$,
$N_{k\alpha}^{(n)}$ is the number of $\alpha$ tile at iteration $n$, and
\begin{widetext}
\begin{equation}
  M_k=\left(
    \begin{array}{cccccc}
      \displaystyle \frac{k(k+1)}{2} & \displaystyle\frac{(k+4)(k-1)}{2} & \displaystyle\frac{k+1}{3} & \displaystyle\frac{k^2+5k-2}{6} & \displaystyle\frac{k(k+1)}{6} & 1 \\ \\
      \displaystyle \frac{(k-1)(k-2)}{2} & \displaystyle\frac{(k-2)(k-3)}{2} & \displaystyle\frac{k-1}{3} & \displaystyle\frac{(k-1)(k-2)}{2} & \displaystyle\frac{(k-1)(k+2)}{2} & 0\\ \\
      3 & 3 & 0 & 1 & 0 & 0 \\ \\ 
      3(k-1) & 3(k-2) & 1 & k-1 & k+1 & 0\\ \\
      0 & 3 & 0 & 1 & 0 & 0 \\ \\
      0 & 0 & 0 & 0 & \displaystyle\frac{1}{3} & 0 
    \end{array}
  \right).
\end{equation}
\end{widetext}
The substitution matrix of the golden-mean tiling
is represented by a $5\times 5$ matrix,
by removing the row and column corresponding
to the B tile from $M_k$ with $k=1$.
The maximum eigenvalue of the matrix is given by $\tau_k^2$ for any $k$. 
Consequently, these modulated honeycomb lattices, derived from the substitution rule, are
characterized by the $k$th metallic-mean.
The tile fractions are exactly obtained from the corresponding eigenvector as
\begin{align}
f_{\rm A}&=c_k(\tau_k^2+4\tau_k+1)\tau_k^2,\\
f_{\rm B}&=c_k(\tau_k^2-\tau_k-1)^2,\\
f_{\rm C}&=3c_k(2\tau_k^2+4\tau_k+1),\\
f_{\rm D}&=3c_k(\tau_k-1)(\tau_k+1)(2\tau_k+1),\\
f_{\rm E}&=3c_k\tau_k^2,\\
f_{\rm F}&=c_k,
\end{align}
where $c_k=(\tau_k+1)^{-4}/2$.
These expressions also hold in the golden-mean case $(k=1)$.
The tile fractions are shown in Fig.~\ref{frac}.
In the golden-mean modulated honeycomb lattice $(k=1)$,
the C tiles account for more than 40 percent of the total,
and no B tiles appear. 
When $k>1$, the A tiles mainly cover the two dimensional sheet. 
Additionally, the B tiles emerge and monotonically increase as $k$ increases.
We find that the fractions of A and B tiles converge to half 
in the limit $k\rightarrow \infty$, where
two dimensional sheet is covered with A and B tiles (see Fig.~\ref{lattice}).
Thus, this class of modulated honeycomb lattices can be regarded as
aperiodic approximants of the honeycomb lattice,
which is distinct from
those proposed recently~\cite{Nakakura_2019,MatsubaraNC,Nakakura_2024}.

We note here that, in the large $k$ case, 
A and B tiles are not homogeneously mixed in the two-dimensional sheet; instead,
they form domain structure consisting of adjacent A or B tiles,
with the domains arranged alternately, as shown in Fig.~\ref{lattice}(d).
We also find a clear difference between A and B domains.
Each A domain is bounded by C, D, and E tiles,
while each B domain is bounded by D tiles.
These originate from the matching rules of the A and B tiles,
which are never adjacent to each other.
Furthermore, we find that, for the $k$th metallic-mean tiling,
the average number of A tiles in the A domain is larger than the other,
which is consistent with $f_{\rm A}> f_{\rm B}$.
This domain property is different from that for the aperiodic approximants
for the honeycomb lattice proposed in the previous study~\cite{MatsubaraNC}, 
where two types of domains are identical due to the presence of a parallelogram among the prototiles.

\begin{figure}[htb]
  \includegraphics[width=\linewidth]{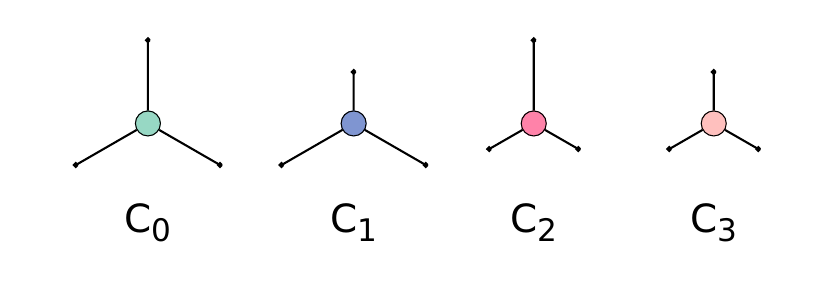}
  \caption{
    Four kinds of vertices in the modulated honeycomb lattices. 
    C$_i$ vertex is connected to the nearest neighbor vertices 
    by $i$ short bonds and $(3-i)$ long bonds.
    The ratio between long and short lengths is $\tau_k$.
    }
  \label{vertex}
\end{figure}
\begin{figure}[htb]
  \includegraphics[width=\linewidth]{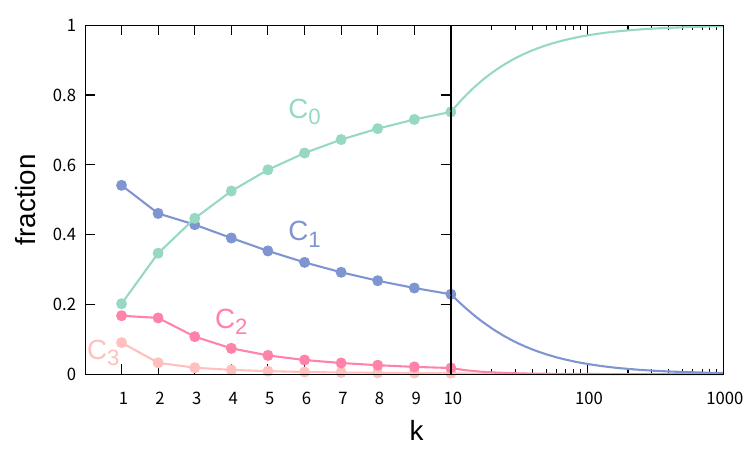}
  \caption{
    Vertex fractions of the $k$th metallic-mean modulated honeycomb lattices.
  }
  \label{vfrac}
\end{figure}

Next, we discuss vertex properties in the modulated honeycomb lattices.
There exist four types of vertices C$_0$, C$_1$, C$_2$, and C$_3$,
where the C$_i$ vertex is connected by $i$ short bonds and $(3-i)$ long bonds,
as shown in Fig.~\ref{vertex}.
By using the substitution rules,
the vertex fractions are derived. 
Exact results are classified into three cases: 
$k=1$, $k=2$, and $k\ge 3$.
The details are provided in Appendix~\ref{vertexf}.
The vertex fractions are shown in Fig.~\ref{vfrac}.
We find that, as $k$ increases, the fraction of the C$_0$ vertices
monotonically increases, while the others decrease.
This is consistent with the fact that, in the large $k$ limit,
the system reduces to the regular honeycomb lattice
where the distance between the nearest neighbor sites is $\ell$.

\begin{figure}[htb]
  \includegraphics[width=0.8\linewidth]{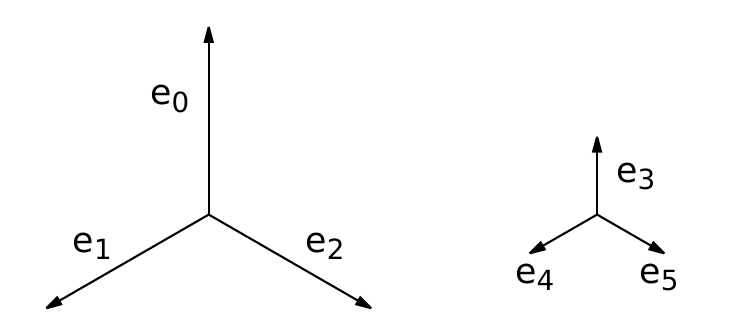}
  \caption{
    Projected basis vectors ${\bf e}_m\;(m=0, \cdots, 5)$
    from fundamental translation vectors in six dimensions.
    The ratio between long and short lengths is $\tau_k$.
  }
  \label{vector}
\end{figure}

When discussing spatial profile of the vertices characteristic of
the metallic-mean tilings,
the perpendicular space analysis is instructive. 
The positions in perpendicular space have one-to-one correspondence
with those in the physical space. 
The vertex sites in the modulated honeycomb lattice are represented by a subset of 
the six-dimensional lattice points $\vec{n}=(n_0, n_1, n_2, n_3, n_4, n_5)$, 
where $n_m$ is an integer.
Their coordinates are the projections 
onto the two-dimensional physical space:
\begin{align}
  {\bf r}=(x,y)=\sum_{m=0}^5n_m {\bf e}_m,
\end{align}
where 
${\bf e}_m= (\ell \cos(m\theta+\theta_0), \ell \sin(m\theta+\theta_0))$ 
for $m=0, 1, 2$, and ${\bf e}_m=(s\cos(m\theta+\theta_0), s\sin(m\theta+\theta_0))$ 
for $m=3, 4, 5$ with $\theta=2\pi/3$ and initial phase $\theta_0=\pi/2$. 
The projected basis vectors ${\bf e}_m$ are shown in Fig.~\ref{vector}.
The projection onto the four-dimensional perpendicular space has information 
specifying the local environment of each site,
\begin{align}
  {\bf \tilde{r}}&=(\tilde{x},\tilde{y})=\sum_{m=0}^5n_m {\bf \tilde{e}}_m,\\
  {\bf r^\perp}&=(x^\perp,y^\perp)=\sum_{m=0}^5n_m {\bf e}^\perp_m,
\end{align}
where ${\bf \tilde{e}}_m={\bf e}_{m+3}$ and ${\bf \tilde{e}}_{m+3}=-{\bf e}_{m}$ $(m=0,1,2)$,
and ${\bf e}_m^\perp=(\delta_{m\; ({\rm mod}\; 2), 0}, \delta_{m\; ({\rm mod}\; 2), 1})$.
${\bf r}^\perp$ takes only six values.
When a certain C$_3$ vertex is appropriately chosen as
the origin in six dimensions,
vertices appear in planes ${\bf r}^\perp =(0,1), (0,0), (1,0), (1,-1), (2,-1)$, and $(2,-2)$.
In each ${\bf r}^\perp$ plane,
the ${\bf \tilde{r}}$ points densely cover a triangular or hexagonal window.

\begin{figure}[htb]
  \includegraphics[width=\linewidth]{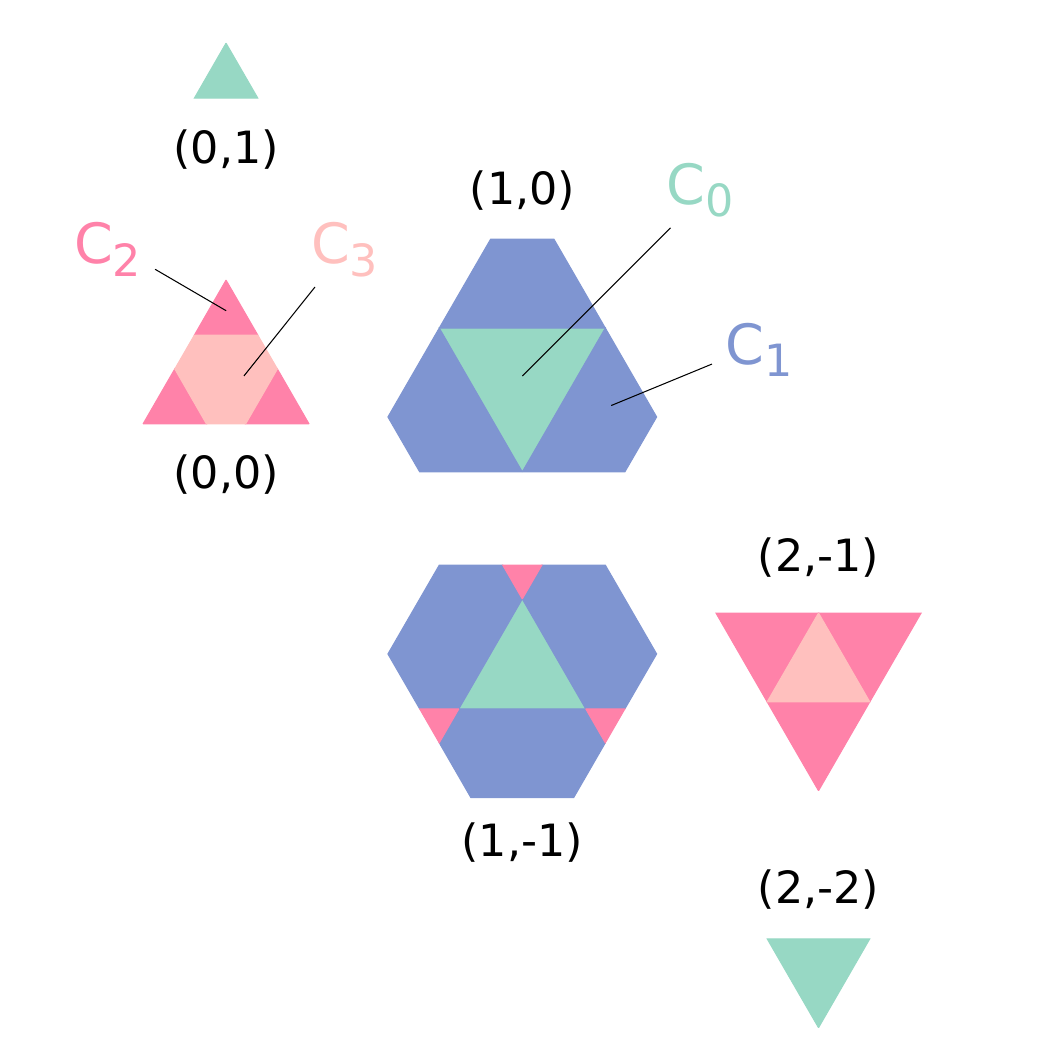}
  \caption{
    Perpendicular space in the golden-mean modulated honeycomb lattice.
    Each part is the window of four types of vertices shown in Fig.~\ref{vertex}.
  }
  \label{perp}
\end{figure}
Figure~\ref{perp} shows the perpendicular spaces in the golden-mean modulated honeycomb lattice,
where the colored windows represent the four types of vertices.
A key characteristic of this structure is that each window in the plane
specified by ${\bf r}^\perp$ has a unique shape.
In addition, each vertex appears in a distinct shaped window on specific planes.
This is in contrast to the well-known bipartite quasiperiodic tilings 
such as Penrose and Socolar-dodecagonal tilings.
In these cases, identical windows appear in pairs across the planes and
their symmetry originates from the equivalence of two sublattices.

Now, we examine sublattice properties of the modulated honeycomb lattice.
Here, the sublattice for the vertex with $\vec{n}=(0,0,0,0,0,0)$ 
is defined as the A sublattice.
Then, the vertices in the planes specified by ${\bf r}^\perp=(0,0), (1,-1), (2,-2)$ belong to the A sublattice, 
whereas the others belong to the B sublattice.
This classification follows from the fact that
moving from one site to its neighboring site changes 
only one component of $\vec{n}$ by $\pm 1$,
which shifts either $x^\perp$ or $y^\perp$ by $\pm 1$.
We find that the sublattice imbalance arises 
when focusing on a certain type of vertices.
The sublattice imbalances for $\alpha$(=C$_0$, C$_1$, C$_2$, C$_3$) vertices
are explicitly given as
\begin{align}
  \Delta_{{\rm C}_0}&=-\frac{1}{2\tau_1^7},\\
  \Delta_{{\rm C}_1}&=\frac{3}{2\tau_1^7},\\
  \Delta_{{\rm C}_2}&=-\frac{3}{2\tau_1^7},\\
  \Delta_{{\rm C}_3}&=\frac{1}{2\tau_1^7},
\end{align}
where $\Delta_\alpha=f_{\alpha,A}-f_{\alpha,B}$ and
$f_{\alpha\sigma}$ is the fraction of the $\alpha$ vertex
in the sublattice $\sigma(=A, B)$.
$\sum_\alpha\Delta_\alpha=0$ means the absence of the sublattice imbalance
when the total vertices are considered.
These results should induce nontrivial magnetic properties 
if one considers antiferromagnetic correlations,
which will be discussed in Sec.~\ref{sec:mag}.
Additional analyses of perpendicular spaces of other metallic-mean tilings
are provided in Appendix~\ref{perpA}.

We have obtained the modulated honeycomb lattices composed of six prototiles
with lengths $\ell$ and $s$.
This structure allows us to construct a vertex model with
two types of couplings~\cite{MatsubaraProc}.
The simplest models we consider are tight-binding and Hubbard models,
whose ground-state properties will be discussed in the following sections.
It is important to note that these models are defined by
the point-to-point connectivity;
thus, one might assume that
analyzing the coordinates of vertices or their perpendicular space may not 
yield meaningful interpretations. 
Nevertheless, the perpendicular space analysis has an advantage 
in discussing magnetic properties inherent in the quasiperiodic tilings
since it systematically captures the effect of local coordinations around the vertices.
Another important aspect of the model is that it can account for
the effects of disorder.
In fact, we can construct the disordered tight-binding and Hubbard models 
where two types of hopping integrals are randomly distributed in a given ratio.
In the quasiperiodic tiling, the number ratio of long and short bonds 
$r_b=N_L/N_S$ is given by the metallic mean $\tau_k$,
where $N_L$ and $N_S$ are the numbers of long and short bonds, respectively.
This formulation enables a direct comparison
between quasiperiodic and disordered systems.
In the following section,
we examine the DOS of the noninteracting system
to discuss the effects of quasiperiodicity and disorder.

\section{Tight-binding model}\label{sec:tb}

In this section, we consider the tight-binding model 
on the modulated honeycomb lattices
to discuss how the quasiperiodic structure in the hopping integrals affects
the low-energy states.
The effects of disorder are also addressed in the end of this section.
The tight-binding Hamiltonian for the metallic-mean modulated honeycomb lattice is given as
\begin{align}
  H=-t_S\sum_{(ij)}\left( c^\dag_ic_j+h.c.\right)-t_L\sum_{\langle ij\rangle}\left(c^\dag_ic_j+h.c.\right),
\label{H}
\end{align}
where $(ij)\; [\langle ij\rangle]$ stands for the nearest-neighbor pair
on the short (long) edges in the tiling.
$c_i (c^\dag_i)$ is the annihilation (creation) operator of the fermion at the $i$th site.
$t_S$ and $t_L$ are the hopping integrals for the short and long edges of the tiles.
Here, we focus on the DOS defined as,
\begin{align}
  \rho(E)&=\frac{1}{N}\sum_i\delta(E-E_i),
\end{align}
where $E_i$ is the $i$th eigenvalue of the Hamiltonian eq.~(\ref{H})
and $N$ is the number of sites.
When $t_S=t_L$, the model eq.~(\ref{H})
reduces to the tight-binding model on the regular honeycomb lattice.
In the system, there exist two features in the DOS.
One of them is the linear DOS around $E=0$, as
\begin{align}
  \rho(E)\sim \lambda \frac{|E|}{t_S^2},
\end{align} 
with $\lambda=(\sqrt{3}\pi)^{-1}$.
This originates from the existence of the Dirac cones at K and K' points in the dispersion relation.
The other is the logarithmic divergence at $E=\pm t_S$ due to the van Hove singularity.
When $t_S\neq t_L$, the quasiperiodic structure is introduced
and the wave number is no longer a good quantum number.
It should be interesting how robust the linear DOS
and van Hove singularity are against the quasiperiodicity
in the hopping integrals.
In the limit $t_S=0$,
the system is decoupled to isolated subsystems 
such as domains, star-shaped systems, and C$_3$ vertices (see Fig.~\ref{lattice}).
Therefore, in the large $r_t(=t_L/t_S)$ case,
the system can be regarded as a weakly-coupled subsystems.

\begin{figure}[htb]
  \includegraphics[width=\linewidth]{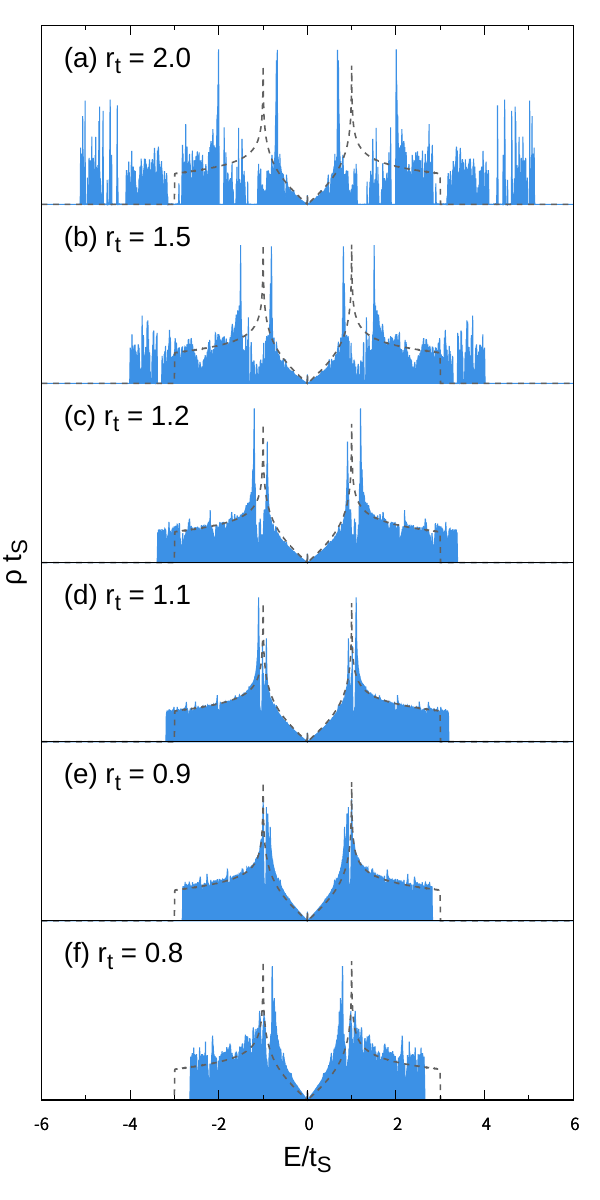}
  \caption{
    DOS for the tight-binding model on the golden-mean
    modulated honeycomb lattices with several $r_t(=t_L/t_S)$ and $N=2,078,532$.
    Dashed lines represent the DOS for the tight-binding model with $r_t=1$.
  }
  \label{dos}
\end{figure}

To analyze the effect of the quasiperiodic structure in the hopping integrals,
we treat the golden-mean modulated honeycomb lattice with open boundary conditions.
We numerically diagonalize the tight-binding Hamiltonian with
$N=2,078,532$
and obtain the DOS, as shown in Fig.~\ref{dos}.
The particle-hole symmetry is clearly found in the DOS
since the system is bipartite.  
We note that there exist 352 zero-energy modes
in the system
even though it does not exhibit the sublattice imbalance.
This is in contrast to the case with periodic boundary conditions
where only four extended states are degenerate at $E=0$.
The existence of the other multiple zero energy modes
originates from the boundary of the system.
This is consistent with the fact that the degeneracy is
independent of the ratio $r_t$
and become negligible with increasing $N$.
Therefore, we conclude the absence of confined states at $E=0$
in this tight-binding model.
This is in contrast to those on
the well-known quasiperiodic tilings 
such as Penrose, Ammann-Beenker, and Socolar dodecagonal 
tilings~\cite{Kohmoto_Sutherland_1986,Arai_1988,Kraj_1988,Koga_Tsunetsugu_2017,Koga_AB,Koga_SC,Oktel_2021,Koga_Coates_2022,Oktel_2022,Keskiner_Oktel_2022,Ghadimi_2023,Matsubara_2024},
where macroscopically degenerate states exist at $E=0$.

\begin{figure}[htb]
  \includegraphics[width=\linewidth]{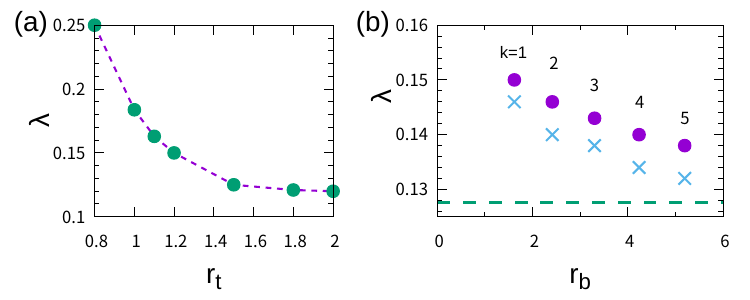}
  \caption{
    (a) Slope $\lambda$ as a function of $r_t(=t_L/t_S)$
    in the system on the golden-mean tiling.
    (b) Circles represent the slope in the system with $r_t=1.2$ 
    on $k$th metallic-mean tiling. 
    Dashed line indicates the slope for the tight-binding model on the regular honeycomb lattice
    realized in $k\rightarrow\infty$.
    Crosses represent the slope in the disordered systems where
    two hopping integrals $t_S$ and $t_L$ are randomly distributed
    according to the ratio $\tau_b(=N_L/N_S)$ (see text).
    }
  \label{slope}
\end{figure}

When $r_t$ is away from unity,
the quasiperiodic structure is introduced in hopping integrals.
We find that the peak structure at $E=\pm t_S$, 
which corresponds to van Hove singularity, is split into two.
This suggests that the macroscopically degenerate states at $E=\pm t_S$
are partially lifted by the quasiperiodic structure.
When $r_t$ is large, the system can be regarded
as a weakly-coupled subsystems,
resulting in multiple energy gaps in the DOS,
as shown in Fig.~\ref{dos}(a).
On the other hand, in the vicinity of $E=0$,
the linear DOS seems to appear for any $r_t$.
This may be explained as follows.
Since our system can be regarded as a continuously deformed honeycomb lattice,
the quasiperiodic structure is introduced
by gradually changing the hopping integrals.
The four degenerate extended states at $E=0$, which exist when $t_S=t_L$, 
still remain even when $t_S\neq t_L$.
Some details are given in Appendix~\ref{sec:E0}.
Therefore, the introduction of the quasiperiodic structure
in the hopping integrals does not lead to a drastic change around $E=0$,
leading to the robust linear DOS.
Figure~\ref{slope}(a) shows the $r_t$ dependence of the slope $\lambda$ 
in the golden-mean modulated honeycomb lattice,
which is roughly deduced from the numerical data.
We find a monotonic decrease in the slope, 
which originates from the fact that 
bonds with the larger hopping integral $t_L$ become dominant.

Similar behavior is also observed in the tight-binding model on
the generic metallic-mean tilings.
\begin{figure}[htb]
  \includegraphics[width=\linewidth]{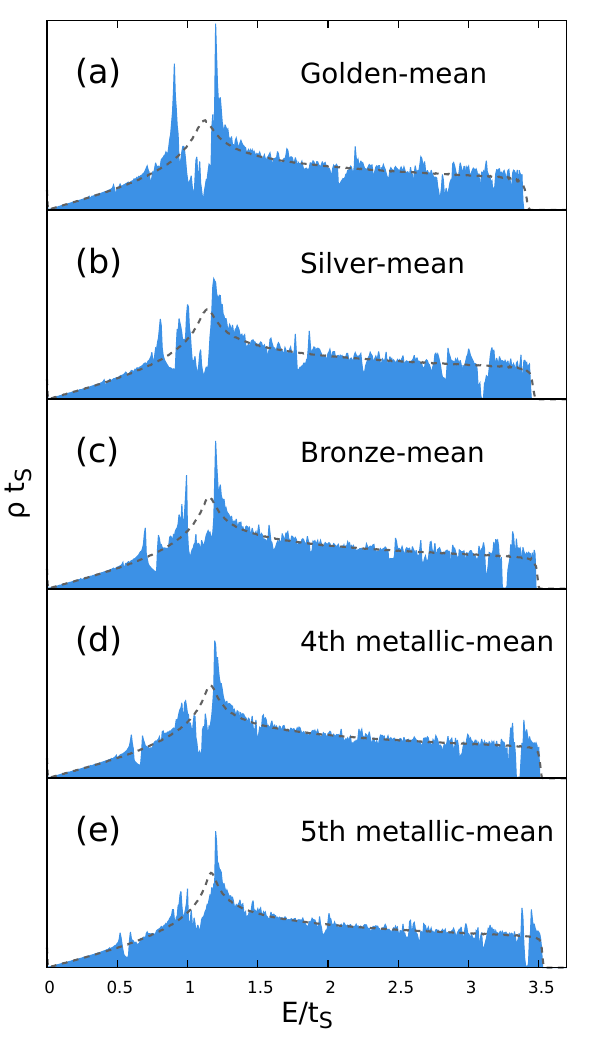}
  \caption{
    DOS for the tight-binding model on (a) golden-mean, (b) silver-mean, 
    (c) bronze-mean, (d) 4th metallic-mean, and (e) 5th metallic-mean
    modulated honeycomb lattices 
    with $N\sim 2,000,000$ when $r_t=1.2$.
    Dashed lines represent the DOS for the tight-binding model
    with randomly-distributed hopping integrals $t_S$ and $t_L$ (see text). 
  }
  \label{dos-add}
\end{figure}
Figure~\ref{dos-add} shows the DOS of the tight-binding model on
the $k$th metallic-mean modulated honeycomb lattices with $k=1, 2, \cdots, 5$,
by diagonalizing the Hamiltonian with $r_t=1.2$ and
$N\sim 2,000,000$.
We find common properties in the DOS.
Spiky peaks and (pseudo)gaps appear in the high energy region
and a linear behavior is observed in the low energy region.
As $k$ increases,
the band edge shifts toward $3t_L$, the peaks at $E=\pm t_L$ sharpens,
and the slope of the DOS approaches $(\sqrt{3}\pi)^{-1} r_t^{-2}$
[see Fig.~\ref{slope}(b)] 
since the two-dimensional sheet is mainly covered with
the A and B tiles with long bonds.
This reflects the structural property of
the aperiodic approximants.

Now, we discuss the effects of quasiperiodicity and disorder
on the DOS.
To this end, we consider the disordered tight-binding model 
where two hopping integrals $t_S$ and $t_L$ are randomly distributed
according to the given ratio $r_b$.
The results for the systems with $N=622,524$
are shown as the dashed lines in Fig.~\ref{dos-add}.
Since the system size is large enough,
the sample dependence in the DOS is hardly visible in this scale.
We find a smooth DOS in the disordered systems, in contrast to that
for the quasiperiodic systems.
In particular, the peak structures at $E=\pm t_S$ are smeared,
and remain only as broad features.
Therefore, the quasiperiodic order is essential for the spiky structure in the DOS.
On the other hand,
the linear DOS remains around $E=0$.
This suggests that
the degeneracy at $E=0$ is never lifted
even by this random distribution in hopping integrals on the honeycomb lattice.
We also find that the slope is slightly smaller than that for
the quasiperiodic systems, as shown in Fig.~\ref{slope}(b).
This implies that the slope is mainly given by the ratio $r_b=N_L/N_S$, 
rather than the bond distribution.

Here, we have discussed the effects of disorder, 
by introducing two types of hopping integrals. 
As a result, we have demonstrated that
low-energy properties remain unchanged,
suggesting that this type of disorder is irrelevant to 
Anderson localization~\cite{Anderson,Motrunich_2002,Mudry_2003}.
Nevertheless, the question of whether the effects of electron correlations in disordered systems 
differ from those in quasiperiodic systems remains nontrivial,
which will be discussed in the following section.

\section{Magnetic properties in the Hubbard model}\label{sec:mag}
In this section, we study the Hubbard model 
to discuss magnetic properties inherent in the modulated honeycomb lattice.
The Hamiltonian is given by $H=H_0+H_1$ with
\begin{align}
  H_0&=-t_S\sum_{(ij)\sigma}\left( c^\dag_{i\sigma}c_{j\sigma}+h.c.\right)
  -t_L\sum_{\langle ij\rangle\sigma}\left(c^\dag_{i\sigma}c_{j\sigma}+h.c.\right), \\
  H_1&=U\sum_i \left(n_{i\uparrow}-\frac{1}{2}\right)\left(n_{i\downarrow}-\frac{1}{2}\right),\label{H1}
\end{align}
where $c_{i\sigma} (c_{i\sigma}^\dag)$ annihilates (creates) an electron with spin $\sigma$ at the $i$th site and 
$n_{i\sigma}=c_{i\sigma}^\dag c_{i\sigma}$. 
$t_S (t_L)$ denotes the hopping integrals on the short (long) bonds and 
$U$ denotes the onsite Coulomb interaction. 
In this study, we focus on the half-filled case
to discuss magnetic properties.
The chemical potential $\mu$ does not shift from $\mu=0$
for any value of $U$ since the system is bipartite.

Magnetic properties in the half-filled Hubbard model
on a periodic bipartite lattice
have been extensively studied.
It is well known that the introduction of the Coulomb interaction
immediately leads to a magnetically ordered state
when a finite noninteracting DOS is present at the Fermi level.
This phenomenon is also observed in quasiperiodic bipartite systems 
although the local magnetization exhibits a highly intricate spatial 
pattern due to the existence of confined states
at $E=0$~\cite{Koga_Tsunetsugu_2017,Koga_AB,Koga_SC,Matsubara_2024,Koga_Coates_2022}.
However, in the Hubbard model on the regular honeycomb lattice, 
the absence of a DOS at the Fermi level stabilizes the semimetallic state against weak interactions.
In the case, a magnetic phase transition occurs
at a finite critical interaction strength 
$U_c$~\cite{Sorella_1992,Furukawa_2001,Feldner_2010,Sorella_2012,Assaad_2013,Raczkowski_2020,Ostmeyer_2020}.
It has been precisely examined by the Monte Carlo method as
$U_c/t_S\sim 3.835$~\cite{Ostmeyer_2021}.
Based on this analogy, a similar phase transition is expected
in the half-filled Hubbard model on the modulated honeycomb lattice
since the semimetallic state with the linear DOS is realized
in the noninteracting case, as discussed in the previous section.

In this study, we employ the Hartree approximation instead
since a larger system size is necessary 
to clarify magnetic properties inherent in quasiperiodic systems. 
Although this method is too simple
to quantitatively determine the critical interaction,
it still captures the key aspects of the phase transition.
In fact, in the case of the regular honeycomb lattice,
the method yields a finite critical interaction, 
although $U_c/t_S\sim 2.23$~\cite{Sorella_1992}
is smaller than the value obtained by the Monte Carlo method.
This mean-field result suggests that
the phase transition can be described by a simplified approach
which takes into account only the characteristic linear DOS.
In this linearized method,
the critical interaction is given by $U_c/t_S=\lambda^{-1/2}$,
with details explicitly presented in Appendix~\ref{sec:linear}.
Notably, in the periodic case, $U_c/t_S\sim 2.33$ is approximately five percent 
larger than the result obtained by the full Hartree approximation.
Therefore, this alternative method allows for qualitative discussions
of the critical interaction.

In the full Hartree approximations,
the Hamiltonian for the Coulomb interactions reduces to
\begin{align}
  H_1&\rightarrow U\sum_{i\sigma}
  \left(\langle n_{i\bar{\sigma}}\rangle-\frac{1}{2}\right) n_{i\sigma},
\end{align}
where the site- and spin-dependent mean-field $\langle n_{i\sigma}\rangle$ 
is the expectation value of the number of electrons with spin $\sigma$ at the $i$th site.
For given mean-field values,
we numerically diagonalize the mean-field Hamiltonian and
update the mean fields, and iterate this self-consistent procedure
until the result converges within numerical accuracy.

\begin{figure}[htb]
  \includegraphics[width=\linewidth]{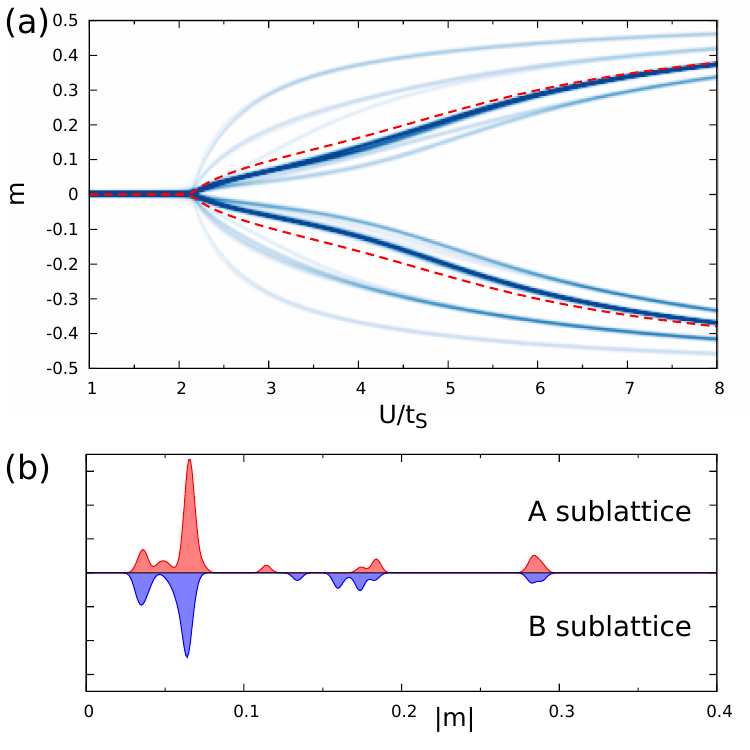}
  \caption{
    (a) Density plot of local magnetizations and (b) its cross section 
    as a function of the Coulomb interaction $U/t_S$
    in the system with $r_t=2$ and $N=98, 005$.
    Dashed lines represent the average of the magnetization
    and a cross indicates
    the critical interaction obtained by the simplified method.
    }
  \label{mag}
\end{figure}
We apply real-space Hartree calculations to the Hubbard model
on the modulated honeycomb lattice with open boundary conditions
to discuss the competition between nonmagnetic semimetallic and magnetically ordered states.
Figure~\ref{mag}(a) shows
the distribution of local magnetizations
for the system with $r_t=2$ and $N=98,005$.
We find that all sites have zero magnetizations when $U<U_c$.
Therefore, in the weak coupling regime,
the system remains in a nonmagnetic state.
A further increase in interaction strength beyond $U_c$ drives the system 
to a magnetically ordered state,
as shown in Fig.~\ref{mag}(a).
We find that the magnetizations splits into several groups
and increase in magnitude.
To clarify this,
we show the cross-section of the distribution for the case with $U/t_S=3$
in Fig.~\ref{mag}(b).  
The distribution is found to separate into multiple peaks.
These groups should be classified by
the local environment of the vertices, which will be discussed below.
One of the remarkable points is the asymmetry in
the magnetization distribution between the A and B sublattices.
This is due to sublattice imbalance for each vertex type,
as discussed in Sec.~\ref{sec:lattice}.
Nevertheless, the total magnetization remains zero. 
This is guaranteed by Lieb's theorem~\cite{Lieb}, which states
the ground state on a bipartite lattice has the total spin
$S_{tot} = 1/2|N_A-N_B|$ 
with $N_A$ and $N_B$ being the total numbers of sites
in A and B sublattices, respectively.

\begin{figure}[htb]
  \includegraphics[width=\linewidth]{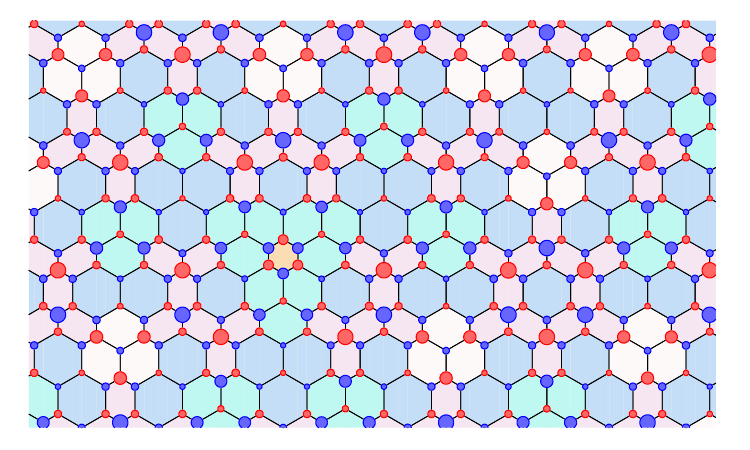}
  \caption{
    Spatial pattern for the staggered magnetization in the Hubbard model 
    on the modulated honeycomb lattice when $U/t_S=3$ and $r_t=2$.
    The area of the circles represents the normalized magnitude of
    the local magnetization.
  }
  \label{mag-phys}
\end{figure}
To clarify in detail how the magnetizations are related to the local environment,
we show in Fig.~\ref{mag-phys} the spatial pattern for the staggered magnetization 
$m_i[=(n_{i\uparrow}-n_{i\downarrow})/2]$ in the half-filled Hubbard model with 
$U/t_S=3$ and $r_t=2$.
The system is bipartite, and the antiferromagnetically ordered state
is clearly found.
Furthermore, we find that the magnitude of the magnetization strongly depends on the site.
Namely, larger magnetizations appear in the C$_i$ vertices with larger $i$.
This can be explained as follows.
If one focuses on a certain vertex C$_i$,
the effective Coulomb interaction may be given by $U/[it_S+(3-i)t_L]$.
Therefore, the magnetizations are spatially distributed,
according to these effective interactions.
This behavior is clearly found in Fig.~\ref{mag}(a) in the large $U$ case,
where the magnetizations are mainly classified into four groups.

\begin{figure}[htb]
  \includegraphics[width=\linewidth]{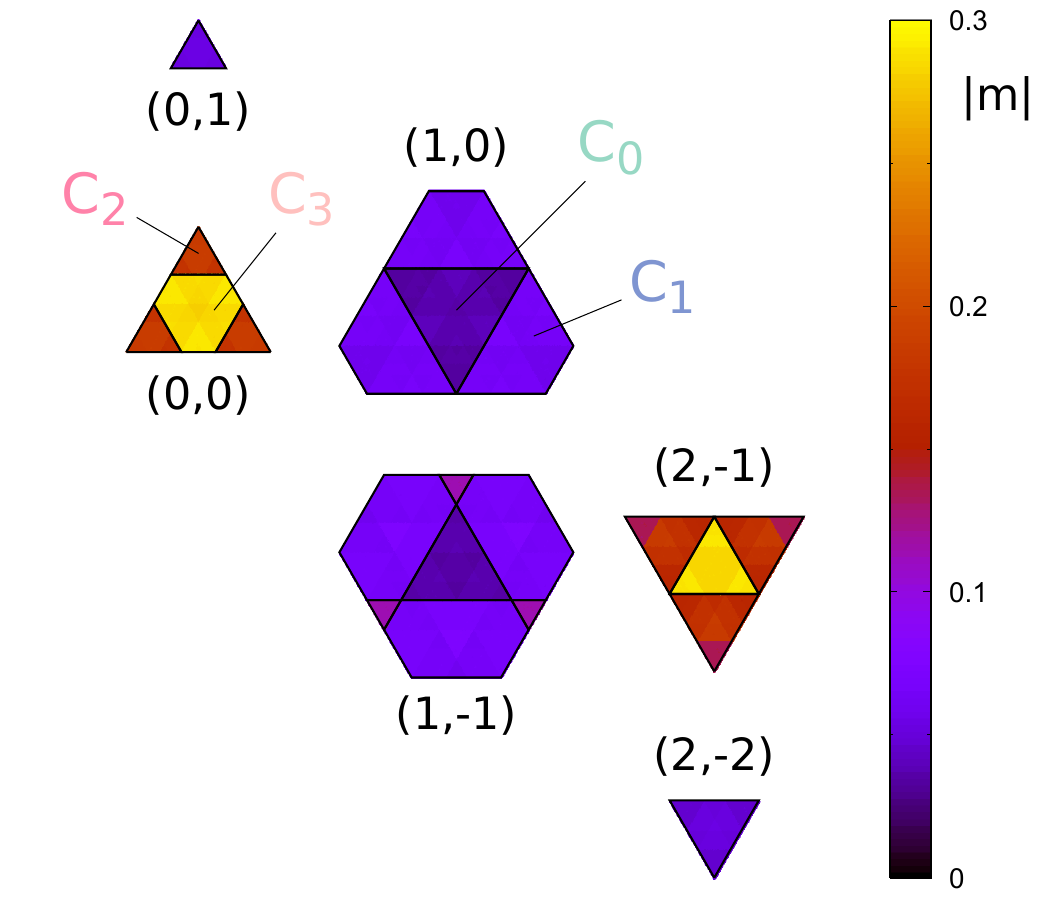}
  \caption{
    Magnetization profile in the perpendicular space for
    the system with $N=350,545$
    when $U/t_S=3$ and $r_t=2$.
  }
  \label{mag-perp}
\end{figure}

To be more precise, we would like to discuss how
the magnetization depends on local environment
around the vertex.
We show in Fig.~\ref{mag-perp}
the magnetization profile in the perpendicular space
when $U/t_S=3$, $r_t=2$, and $N=350,545$.
In this calculation, to focus on bulk magnetic properties,
we discard the boundary sites
located within approximately five units
from the edge of the circular system and map the local magnetization
in the bulk to the perpendicular space.
We find that local magnetizations are roughly classified
according to the vertex type, as discussed above.
Each window consists of subtly different colored subpatterns,
which are smaller triangles and trapezoids.
This reflects differences in the environment not only at the vertex level
but also across larger spatial scales.
Such fine magnetic structures in the perpendicular space 
have also been found in the Hubbard and Heisenberg models 
on some quasiperiodic tilings~\cite{Jagannathan_2007,Koga_Tsunetsugu_2017,Koga_AB,Koga_SC,Matsubara_2024,Koga_Coates_2022}.

Next, we discuss the interaction dependence of the magnetizations,
as shown in Fig.~\ref{mag}(a).
When $U>U_c$,
the curve corresponding to the group with small magnetizations
exhibits shoulder-like behavior around $U/t_S\sim 4$,
while the group with larger magnetizations increases monotonically.
This complex behavior arises from the presence of distinct local structures,
as discussed above.
As the interaction strength decreases from the strong coupling regime,
four curves, initially classified by vertex type, 
further split into multiple branches, 
suggesting that the local environment beyond the vertex plays an important role 
for the magnetization profile.
Finally, a unique phase transition occurs at $U_c/t_S\sim 2.2$.
As a consequence, the average magnetization curve exhibits shoulder-like behavior, 
which is shown as the dashed line in Fig.~\ref{mag}(a).
This is in contrast to magnetic properties in 
the Hubbard model on the regular honeycomb lattice~\cite{Raczkowski_2020}.
A similar magnetic profile is also expected in the Hubbard model
on the other metallic-mean tilings.
Some results for the silver-mean tiling are explicitly shown in Appendix~\ref{sec:silver}.

\begin{figure}[htb]
  \includegraphics[width=\linewidth]{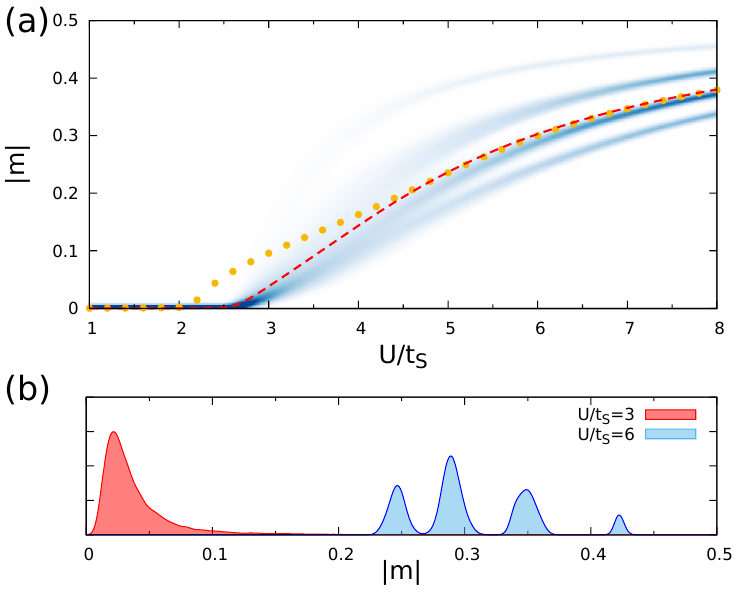}
  \caption{
    (a) Density plot of local magnetizations 
    as a function of the Coulomb interaction $U/t_S$
    in the disordered system with $r_t=2$, $r_b=\tau_1$ and $N=32, 400$.
    Dashed line represents the average magnetization of the disordered system
    and a cross indicates the critical interaction obtained by the simplified method.
    (b) Cross sections of the distribution of magnetization
    at $U/t_S=3$ and $6$. 
    Dotted line in (a) represents the average magnetization of 
    the golden-mean modulated honeycomb system, for reference (see text).
    }
  \label{disorder}
\end{figure}

Here, we clarify the difference in magnetic properties
between the quasiperiodic and disordered systems.
To this end, we also apply the real-space Hartree mean-field approach 
to the disordered system with $r_t=2$ and $r_b=\tau_1$.
Since two types of hopping integrals are randomly distributed,
no sublattice asymmetry is expected in the thermodynamic limit $N\rightarrow\infty$.
The magnetic profile for the disordered system with $N=32,400$
is shown in Fig.~\ref{disorder}(a).
In the strong coupling regime, 
the local magnetizations are classified into four groups, 
as clearly observed in the cross-section for the case $U/t_S=6$ [see Fig.~\ref{disorder}(b)].
This behavior is essentially the same as that observed in the golden-mean modulated honeycomb lattice.
As the interaction strength decreases, 
the distribution of the magnetization broadens,
and a single broad peak appears around $U/t_S\sim 3$, as shown in Fig.~\ref{disorder}(b).
This broadening arises from the random distribution of hopping integrals. 
The phase transition occurs to the semimetallic state at $U/t_S\sim 2.6$.

Now, we compare the average magnetization curves for the quasiperiodic and disordered systems.
We find that both curves are almost identical in the strong coupling regime,
where the magnetic profile depends on the vertex type.
By contrast, distinct behavior emerges near $U/t_S\sim 3$,
where the average magnetization for the quasiperiodic case is larger than the other.
In this parameter regime, multiple branches in the magnetization profile emerge
in the quasiperiodic system, while a broad peak appears in the disordered system.
This suggests that electron correlation and quasiperiodic modulation of the hopping integrals,
which leads to spatially ordered local magnetic structures,
play significant roles in stabilizing
the antiferromagnetically ordered state.

\begin{figure}[htb]
  \includegraphics[width=\linewidth]{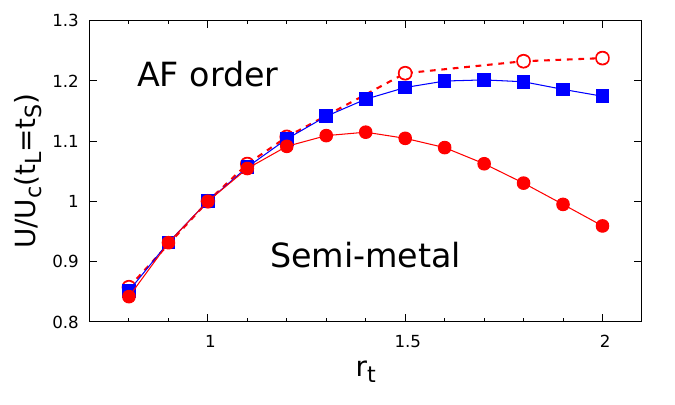}
  \caption{
    Phase diagram of the Hubbard model on the golden-mean
    modulated honeycomb lattice.
    Solid (open) circles represent the phase boundary
    between the magnetically ordered and semimetallic states, which are 
    obtained using the full Hartree approach and linearized method.
    Solid squares represent the phase boundary in the Hubbard model 
    on the disordered honeycomb lattice with $r_b=\tau_1$.
    }
  \label{PD}
\end{figure}
By performing similar Hartree mean-field calculations
for several values of $r_t$,
we obtain the phase diagram of the Hubbard model on the modulated honeycomb lattice,
as shown in Fig.~\ref{PD}.
In the vicinity of the transition point, where the magnetization remains small, 
the results are sensitive to the system size and boundary conditions. 
Accordingly, the transition point is determined 
by extrapolating from the behavior of the magnetization at slightly larger values.
When $r_t=1$, the system corresponds to the Hubbard model on the regular honeycomb lattice,
where the critical interaction strength is $U_c/t_S=2.23$~\cite{Sorella_1992}.
Away from $r_t=1$, the phase boundary shifts upward
since the energy scale also increases accordingly.
This behavior is consistent with that obtained by the linearized approach,
which is shown as the dashed line with open circles in Fig.~\ref{PD}.
This suggests that the slope of the DOS plays an essential role
in the phase transition
between semimetallic and insulating states.
As $r_t$ further increases, 
the critical interaction reaches a maximum around $r_t\sim 1.4$ and
then decreases, 
in contrast to
the rough estimate from the linearized approach.
This discrepancy indicates that
the quasiperiodic structure,
which is not taken into account in the linearized approach,
likely plays an important role
in determining the critical interaction far from $r_t=1$.
Additional support for this interpretation comes from the Hubbard model on the disordered honeycomb lattice,
whose results are shown as squares in Fig.~\ref{PD}.
In this case, the lack of quasiperiodic order does not stabilize the antiferromagnetically ordered state,
which is consistent with the prediction of the linearized approach.

In this analysis for the magnetic properties, 
we have employed the site-dependent mean-field approximation.
It is known that the mean-field approach tends to overestimate the effect of the Coulomb interaction.
For example, in the Hubbard model on the regular honeycomb lattice, 
the critical interaction $U_c/t_S\sim 3.835$ obtained by the Monte Carlo simulations~\cite{Ostmeyer_2021}, 
while mean-field theory gives a lower estimate of $U_c/t_S\sim 2.23$, as mentioned before.
By analogy, we expect that the true critical interaction in our system 
is approximately twice as large as the mean-field estimate.
Furthermore, in the strong coupling limit, 
the system reduces to the Heisenberg model on the quasiperiodic tilings
with nearest-neighbor exchange couplings $J_S=4t_S^2/U$ and $J_L=4t_L^2/U$.
In the mean-field approximation, the ground state exhibits 
a uniform staggered moment $m_j=\pm 1/2$ at $U\rightarrow\infty$.
However, this differs from the predictions of spin wave theory for the Heisenberg model
where site-dependent reduction appears in magnetic moments.
This reduction arises from inhomogeneous quantum fluctuations, 
mainly determined by the coordination number of each site. 
Since this effect is not correctly captured in the mean-field approximation for the Hubbard model, 
more sophisticated approaches such as the random-phase approximation
would be required for a more accurate description. 
However, such refinements are beyond the scope of the present study. 
Despite this limitation, 
our mean-field analysis successfully captures key magnetic properties in the quasiperiodic systems.

\section{Summary}\label{sec:summmary}

We have proposed modulated honeycomb lattices characterized by the metallic mean,
which consist of six distinct hexagonal prototiles with two edge lengths.
The structural properties have been examined through their substitution rules. 
We have constructed a tight-binding model on the tilings,
introducing two types of hopping integrals
corresponding to the two edge lengths.
By diagonalizing the Hamiltonian on these quasiperiodic tilings, 
we have computed the DOS. 
It has been found that,
the introduction of quasiperiodicity in hopping integrals
induces a spiky structure in the DOS at higher energies, while
the linear DOS at low energies remains robust. 
This contrasts with the smooth DOS in the disordered tight-binding model,
where two types of hopping integrals are randomly distributed
according to the metallic mean.

We have also examined the magnetic properties of
the Hubbard model on modulated honeycomb lattices
by means of real-space Hartree approximations.
Our results show that a magnetic phase transition occurs
at a finite interaction strength
since no DOS appears at the Fermi level.
When $t_L\sim t_S$,
the phase transition point is primarily governed by the linear DOS.
However, far from $t_L = t_S$,
the quasiperiodic structure plays a significant role
in reducing the critical interaction strength.
This behavior arises from the introduction of the modulated lattice as a continuous deformation of the honeycomb lattice. 
We have further analyzed the magnetic profiles in perpendicular space, 
revealing additional signatures of the underlying quasiperiodic geometry.
These findings highlight the rich interplay between geometry and electron correlations in quasiperiodic systems and 
demonstrate that modulated honeycomb lattices provide a valuable platform for exploring emergent quantum phenomena.

\begin{acknowledgments}
  We would like to thank T. Dotera for fruitful discussions.
  Parts of the numerical calculations were performed
  in the supercomputing systems in ISSP, the University of Tokyo.
  This work was supported by Grant-in-Aid for Scientific Research from
  JSPS, KAKENHI Grant Nos. JP22K03525, JP25H01521, JP25H01398 (A.K.),
  and JST SPRING Grants Nos. JPMJSP2106 and No. JPMJSP2180 (T.M).
\end{acknowledgments}

\appendix

\section{Substitution rule for the modulated honeycomb lattices}\label{sub}

Here, we describe how to construct the substitution rule of the modulated honeycomb lattice
using six types of directed tiles scaled by $\tau_k^{-1}$. 
First, we discuss tile properties derived from the matching rule, as shown in Fig.~\ref{6tiles}.
An A (B) tile has a single (double) triangle on each edge and 
some adjacent A (B) tiles can form the honeycomb domains (see Fig.~\ref{lattice}).
In the other words, the A and B tiles are never adjacent.
When we focus on the C$_1$ (C$_0$) vertex, two (three) long edges has a common property:
they contain the same number of triangles and their directions point either toward or away from the vertex.
Therefore, D and E tiles are never adjacent by sharing the short edge, but only the same type of tiles are connected.

\begin{figure}[htb]
  \includegraphics[width=\linewidth]{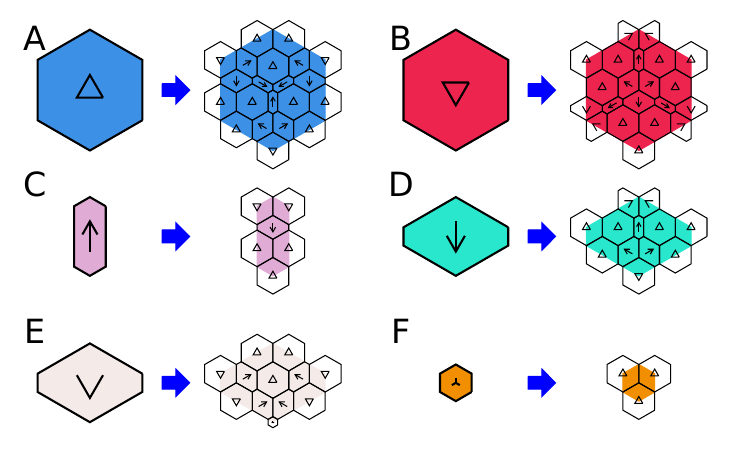}
  \caption{
    Substitution rule for the bronze-mean modulated honeycomb lattice.
  }
  \label{bronze}
\end{figure}

Now, we consider the substitution rule for an A tile.
When an A tile is substituted,
three C tiles are defined to be placed at the center, oriented such that
their directions point toward the center.
Subsequently, some A tiles are uniquely arranged adjacent to these C tiles. 
Additionally, some B and D tiles are arranged so that the matching rule is satisfied.
The substitution rule of the A tile is then obtained, where
it is replaced to $k(k+1)/2$ A tiles, $(k-1)(k-2)/2$ B tiles, 3 C tiles, and $3(k-1)$ D tiles.
By taking into account the matching rule, the substitution rule for the C tile is uniquely determined
since its long edge is identical to that of the A tile.
The same rule is applied to the lower (upper) edges for the D (E) tile. 
One C tile, oriented such that its direction points upward,
is defined to be placed around the center
when a D tile is substituted.
In the case, the substitution rule for the D tile is uniquely determined.
When an E tile is substituted, an F tile appears near the bottom corner.
Then, the substitution rule for the E tile is determined, 
by taking into account the matching rule.
Consequently, the substitution rules for the other tiles are also uniquely determined. 
The substitution rule for the bronze-mean tiling is explicitly shown in Fig.~\ref{bronze}.

\section{Vertex fraction}\label{vertexf}

In this section, we derive the exact vertex fractions, using
the substitution rule.
Here, we focus on the case with $k\ge 3$.
By taking into account the substitution rule,
we find that one C$_3$ vertex only appear in the center 
when the A tile is substituted,
as shown in Fig.~\ref{bronze}.
Therefore, the fraction of the C$_3$ vertex is given as
\begin{align}
  f_{{\rm C}_3}&=R\frac{f_A}{\tau_k^2}=\frac{\tau_k^2+4\tau_k+1}{4(1+\tau_k)^4},
\end{align}
where $R=1/2$ is the number ratio between vertices and tiles.
In the metallic-mean tilings,
the C$_2$ vertices appear both at the corner of the F tiles and away from them,
as shown in Fig.~\ref{lattice}.
The fraction at the corner of the F tiles is given as $6Rf_F$,
while the fraction away from them is given as
$R(3f_A+6f_B+2f_D)/\tau_k^2$,
by taking into account the substitution rules
for the A, B, and D tiles.
As for the C$_1$ vertex,
we find that two appear around each fat (D and E) tile,
six around each F tile, and
three around each C$_3$ vertex.
Therefore, its fraction is given as
\begin{align}
  f_{{\rm C}_1}&=R\left[2(f_C+f_D)+6f_F\right]+3f_{{\rm C}_3}.
\end{align}
The fraction of the C$_0$ vertex is the reminder.

The expressions of the vertex fractions in the golden-mean and silver-mean tilings
are different from the above ones.
For example, in the substitution rule of the silver-mean tilings,
shown in Fig.~\ref{silver},
the C$_3$ vertices are generated when both A and B tiles are substituted,
in contrast to the case with $k\ge 3$ discussed above.
By carefully considering the substitution rule, 
we obtain the exact vertex fractions.
The results are summarized in Table~\ref{tab:vertex}.

\begin{table}
  \caption{Vertex fractions for the metallic-mean modulated
    honeycomb lattices. $c_v=(\tau_k+1)^{-4}/4$.}
  \begin{tabular}{cccl}
    \hline \hline
    Type & $k=1$ & $k=2$ & $k\ge 3$ \\
    \hline
    C$_0$ & $\displaystyle \frac{\sqrt{5}}{\tau_1^5}$ & $\displaystyle \frac{3}{16\tau_2^2}$ & $c_v(4\tau_k^4+4\tau_k^3-\tau_k^2+3)$\\
    C$_1$ & $\displaystyle \frac{6}{\tau_1^5}$ & $\displaystyle \frac{15}{16\tau_2^2}$ & $3c_v(4\tau_k^3+5\tau_k^2-3)$ \\
    C$_2$ & $\displaystyle \frac{3}{\tau_1^6}$ & $\displaystyle \frac{3}{4}-\frac{27}{16\tau_2^2}$ & $3c_v(3\tau_k^2+4\tau_k+3)$ \\
    C$_3$ & $\displaystyle \frac{1}{\tau_1^5}$ & $\displaystyle \frac{1}{4}+\frac{9}{16\tau_2^2}$ & $c_v(\tau_k^2+4\tau_k+1)$ \\
    \hline \hline
  \end{tabular}
  \label{tab:vertex}
\end{table}

\section{Perpendicular spaces for the metallic-mean modulated honeycomb lattice}\label{perpA}

\begin{figure*}[htb]
  \includegraphics[width=\linewidth]{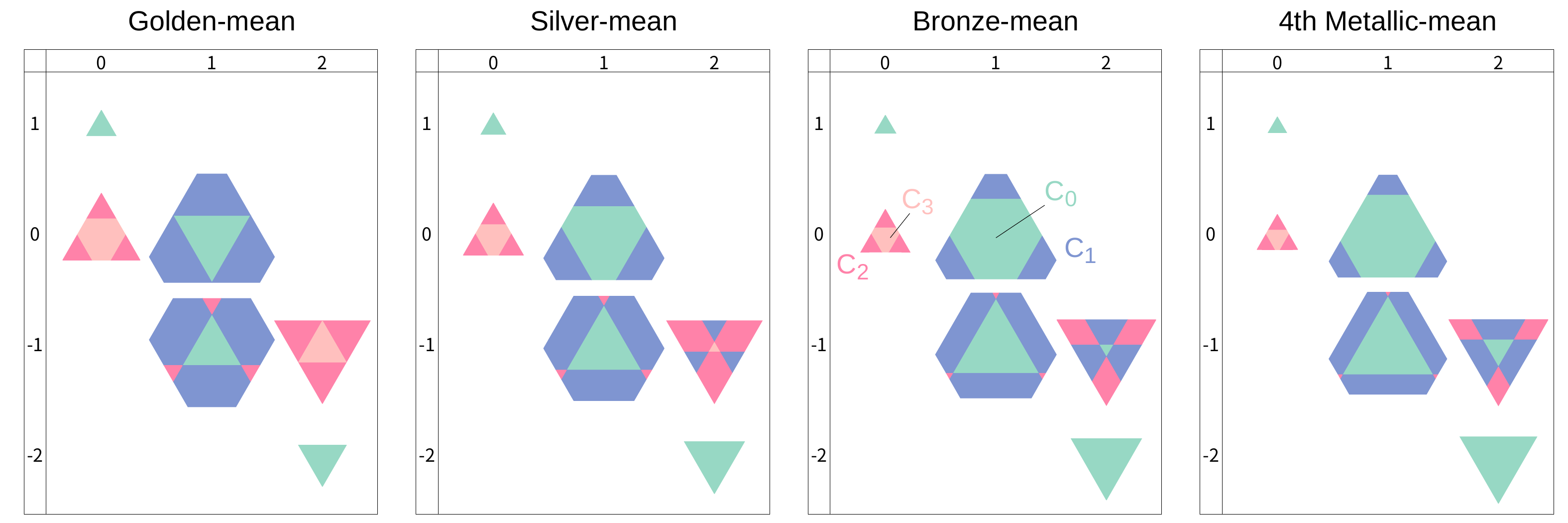}
  \caption{
    Perpendicular space in the modulated honeycomb lattices.
    Each part is the window of four types of vertices
    shown in Fig.~\ref{vertex}.
  }
  \label{perp-silver}
\end{figure*}

The perpendicular space in the metallic-mean modulated honeycomb lattice is shown in Fig.~\ref{perp-silver}.
The area of each colored window is proportional to the corresponding vertex fraction
shown in Table~\ref{tab:vertex}.
As for the silver-mean tiling,
in the plane with ${\bf r}^\perp=(2,-1)$, the vertices C$_1$, C$_2$, and C$_3$ are present,
which differs from the golden-mean tiling, where only C$_2$, and C$_3$ vertices appear.
C$_3$ vertices appear in the planes ${\bf r}^\perp=(0,0)$ and $(2,-1)$
for the golden-mean and silver-mean tilings,
while they appear only in the plane ${\bf r}^\perp=(0,0)$ in the other tilings.
These observations indicate the presence of sublattice imbalance
in the distribution of each vertex type across this family of modulated honeycomb lattices.
Specifically, in the silver-mean modulated honeycomb lattice,
the sublattice imbalances for $\alpha(=$C$_0$, C$_1$, C$_2$, C$_3$) vertices are explicitly given as
\begin{align}
  \Delta_{{\rm C}_0}&=\frac{1}{16}(18\sqrt{2}-25),\\
  \Delta_{{\rm C}_1}&=\frac{1}{16}(18\sqrt{2}-25),\\
  \Delta_{{\rm C}_2}&=-\frac{3}{16}(18\sqrt{2}-25),\\
  \Delta_{{\rm C}_3}&=\frac{1}{16}(18\sqrt{2}-25).
\end{align}

\section{Extended states with $E=0$ in the modulated honeycomb lattice}\label{sec:E0}

\begin{figure}[htb]
  \includegraphics[width=\linewidth]{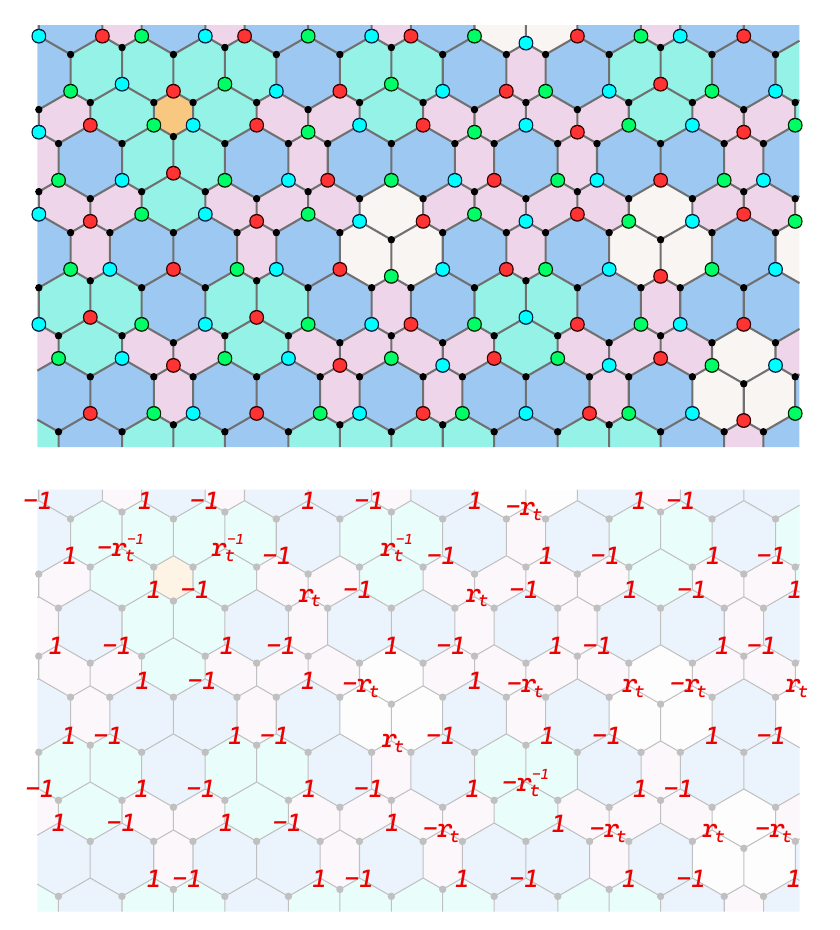}
  \caption{
    Upper panel shows the golden-mean modulated honeycomb lattice.
    Open circles in red, blue, and green represent
    the vertices of three distinct groups within sublattice A,
    while solid circles represent the vertices belonging to sublattice B.
    Lower panel displays the eigenstate $|\Psi_R\rangle$ at energy $E = 0$.
    The values at the vertices represent the amplitudes of $|\Psi_R\rangle$
    with $r_t=t_L/t_S$ (see text).
    }
  \label{E0}
\end{figure}

Here, we consider the degenerate states at $E=0$
in the tight-binding model on the modulated honeycomb lattice.
Before discussing their wave functions,
we first describe the structural properties of the lattice. 
The modulated honeycomb lattice is bipartite and
the vertices in each sublattice are further
classified into three distinct groups.
The upper panel of Fig.~\ref{E0} shows that
the sublattice A can be divided into three groups (R, G, B),
which are shown as open circles in red, green, and blue.
Note that this classification is not clearly visible
in the perpendicular space,
where vertices in each group are uniformly distributed
in the corresponding windows.
This reflects the connectivity
within the honeycomb lattice
rather than the geometry of the quasiperiodic structure.

We now consider the wave functions at $E = 0$
for the tight-binding model on the modulated honeycomb lattice.
Since $H|\Psi\rangle =0$, we can choose a wave function with nonzero amplitude
in one of the sublattices.
The Schr\"odinger equation for the wave function with 
nonzero amplitude in sublattice A is then given by
\begin{align}
  t_{i,i_R} \langle i_R|\Psi\rangle+t_{i,i_G} \langle i_G|\Psi\rangle+t_{i,i_B} \langle i_B|\Psi\rangle= 0,\label{eq:Sch}
\end{align}
where $i$ is the vertex in the sublattice B and
$i_\alpha$\; ($\alpha$=R, G, B) is its nearest neighbor vertex in the $\alpha$th group and
$t_{i,i_\alpha}$ is the hopping integral between $i$ and $i_\alpha$ vertices.
Here, we focus on the wave function $|\Psi_R\rangle$ whose amplitudes are nonzero only on the vertices in the B and G
groups and vanish on all other vertices.
Solving this equation eq.~(\ref{eq:Sch}) uniquely determines the wave function $|\Psi_R\rangle$,
which is explicitly shown in the lower panel of Fig.~\ref{E0}.
We find that the amplitudes take values of
$\pm 1$, $\pm r_t$, and $\pm r_t^{-1}$.
This result can also be confirmed in the perpendicular space,
where $\pm r_t$, $\pm 1$, and $\pm r_t^{-1}$ appear
in the planes ${\bf r}^\perp=(0,0), (1,-1)$, and $(2,-2)$, respectively (not shown).
Therefore, these wave functions at $E=0$ are extended,
which reflects the underlying quasiperiodic structure.
Although the wave functions $|\Psi_G\rangle$ and $|\Psi_B\rangle$,
which have zero amplitudes on the vertices in the G and B groups,
respectively,
can also be constructed,
the three wave functions $(|\Psi_R\rangle, |\Psi_G\rangle, |\Psi_B\rangle)$
are not linearly independent.
Thus, we obtain only two degenerate states associated with sublattice A.
This degeneracy should be closely related to valley degrees of freedom
in the regular honeycomb system
since these wave functions reduce to those of the tight-binding model
on the regular honeycomb lattice in the limit $r_t\rightarrow 1$.
Two additional degenerate states arise from sublattice B.
Therefore, we conclude that
four degenerate extended states at $E=0$, which exist at $t_L=t_S$,
remain robust even when $t_L\neq t_S$.

\section{Critical interaction for the linearized DOS}\label{sec:linear}
\begin{figure}[htb]
  \includegraphics[width=\linewidth]{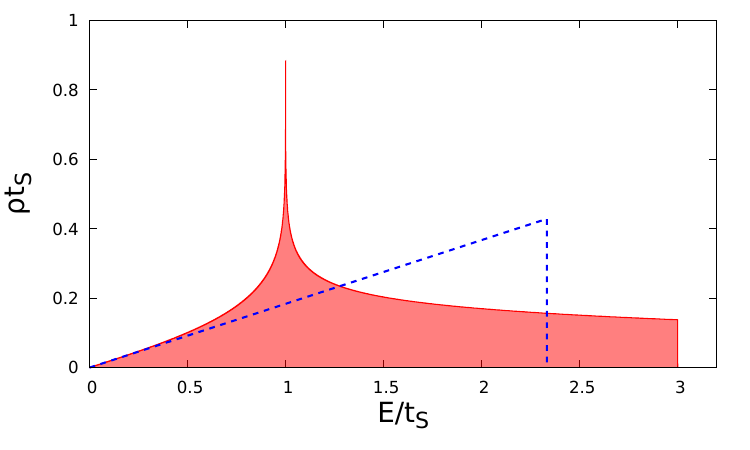}
  \caption{
    Solid line represents the DOS of the tight-binding model 
    on the honeycomb lattice and dashed line represents the linearized DOS.
  }
  \label{linear}
\end{figure}

In the section, we derive the explicit expression
of the critical interaction, using the linearized DOS. 
Considering the DOS of the tight-binding model
on the regular honeycomb lattice,
we can introduce the linearized DOS~\cite{Sorella_1992} as
\begin{align}
  \rho(E)= \lambda \frac{|E|}{t_S^2}\;\; (|E|\le\Lambda),
\end{align}
where the energy cutoff $\Lambda(=\lambda^{-1/2}t_S)$ has been introduced 
to satisfy $\int \rho(E) dE = 1$.
The linearized DOS is shown in Fig.~\ref{linear}.
Since the self-consistency condition is given as
\begin{align}
  \frac{1}{U}&=\frac{1}{2}\int_{-\infty}^\infty\frac{\rho(E)}{\sqrt{E^2+\Delta^2}}dE
  =\frac{\lambda}{t_S^2}\left(\sqrt{\Lambda^2+\Delta^2}-\Delta\right),
\end{align}
where $\Delta=mU$ and $m$ is the staggered magnetization of
the antiferromagnetically ordered state.
Since $\Delta\rightarrow 0$ in the vicinity of the transition point,
we obtain the critical interaction as
\begin{align}
  U_c=\lambda^{-1/2}t_S.\label{uc}
\end{align}
In the case of the Hubbard model on the regular honeycomb lattice,
the critical interaction is given by $U_c/t_S=3^{1/4}\pi^{1/2}\sim 2.33$,
which is comparable to the value obtained using the full Hartree approximation,
$U_c/t_S\sim 2.23$,
as mentioned in the main text.
Therefore, we expect that
the approximation yields
a reasonable estimate of the critical interaction
even in quasiperiodic and disordered systems
although it neglects the spatial dependence of the magnetization.

\section{Magnetic properties in the Hubbard model on the silver-mean modulated honeycomb lattice}
\label{sec:silver}
\begin{figure}[htb]
  \includegraphics[width=\linewidth]{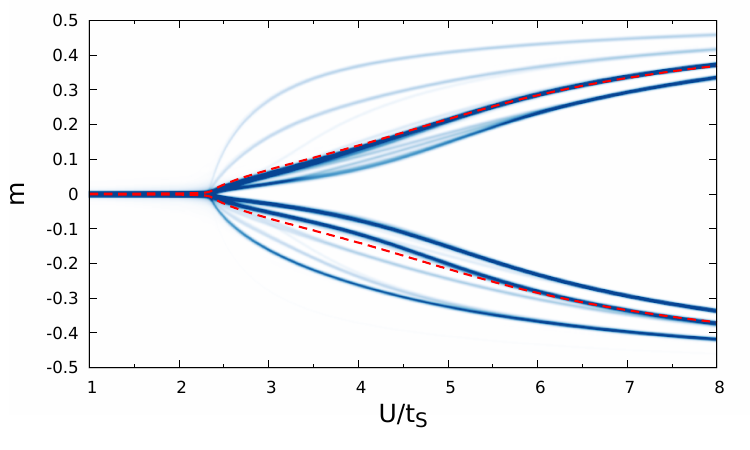}
  \caption{
    Density plot of local magnetizations
    as a function of the Coulomb interaction $U/t_S$ 
    in the Hubbard model on the silver-mean modulated honeycomb lattice
    when $r_t=2$ and $N=96, 546$.
  }
  \label{mag-silver}
\end{figure}

We apply the real-space Hartree approximation to the Hubbard model
on the silver-mean modulated honeycomb lattice
with $r_t=2$ and $N=96,546$.
The resulting magnetic profile is shown in Fig.~\ref{mag-silver}.
Compared with the results for the golden-mean case discussed in the main text,
the weight of the group with small magnetizations is increased.
This tendency can be understood by the following.
As $k$ increases,
the system is regarded as the coupled large A and B domains.
In the case, magnetic properties in the domains are dominant
even though a single magnetic phase transition occurs.
This leads to the pronounced shoulder-like behavior in the magnetization curve.

\nocite{apsrev42Control}
\bibliographystyle{apsrev4-2}
\bibliography{./refs}
  
\end{document}